\documentclass[12pt,preprint]{aastex}

\begin{document}

\title{SOLAR NEUTRON EVENT IN ASSOCIATION WITH A LARGE SOLAR FLARE ON NOVEMBER 24, 2000}

\author{K. WATANABE\altaffilmark{1}, Y. MURAKI\altaffilmark{1}, Y. MATSUBARA\altaffilmark{1}, K. MURAKAMI\altaffilmark{1}, T. SAKO\altaffilmark{1}, \\ H. TSUCHIYA\altaffilmark{2}, S. MASUDA\altaffilmark{1}, M. YOSHIMORI\altaffilmark{3}, N. OHMORI\altaffilmark{4}, P. MIRANDA\altaffilmark{5}, \\ N. MARTINIC\altaffilmark{5}, R. TICONA\altaffilmark{5}, A. VELARDE\altaffilmark{5}, F. KAKIMOTO\altaffilmark{6}, S. OGIO\altaffilmark{6}, \\ Y. TSUNESADA\altaffilmark{6}, H. TOKUNO\altaffilmark{6}, Y. SHIRASAKI\altaffilmark{7}}

\altaffiltext{1}{Solar-Terrestrial Environment Laboratory, Nagoya University, Nagoya, 464-8601, Japan}
\altaffiltext{2}{Institute for Cosmic Ray Research, University of Tokyo, Kashiwa, 277-8582, Japan}
\altaffiltext{3}{Department of Physics, Rikkoyo University, Toshima-ku, Tokyo, 171-8501, Japan}
\altaffiltext{4}{Department of Physics, Kochi University, Kochi, Japan}
\altaffiltext{5}{Instituto Investigaciones Fisicas, Universidad Mayor de San Andr\'{e}s, La Paz, Bolivia}
\altaffiltext{6}{Department of Physics, Tokyo Institute of Technology, Meguro-ku, Tokyo, 152-8551, Japan}
\altaffiltext{7}{Institute of Physical and Chemical Research, Wako, 351-0198, Japan}

\begin{abstract}
Solar neutrons have been detected using the neutron monitor located at Mt.\,Chacaltaya, Bolivia, in association with a large solar flare on November 24, 2000.
This is the first detection of solar neutrons by the neutron monitor that have been reported so far in solar cycle 23. 
The statistical significance of the detection is $ 5.5 \,\sigma $.

In this flare, the intense emission of hard X-rays and $ \gamma $-rays has been observed by the {\it Yohkoh} Hard X-ray Telescope (HXT) and Gamma Ray Spectrometer (GRS), respectively.
The production time of solar neutrons is better correlated with those of hard X-rays and $ \gamma $-rays than with the production time of soft X-rays.
The observations of the solar neutrons on the ground have been limited to solar flares with soft X-ray class greater than X8 in former solar cycles.
In this cycle, however, neutrons were detected associated with an X2.3 solar flare on November 24, 2000.
This is the first report of the detection of solar neutrons on the ground associated with a solar flare with its X-ray class smaller than X8. 
\end{abstract}


\section{INTRODUCTION}
\label{introduction}

Since the discovery of cosmic rays, the acceleration mechanism of cosmic rays has been one of the most important themes in cosmic ray research.
Most cosmic rays are ions, such as protons.
Charged particles are reflected by magnetic fields at the acceleration site, and by interstellar, and interplanetary magnetic fields.
When they arrive at the Earth, exact information on the acceleration place and acceleration time have already been lost.
On the other hand, neutral particles, such as $ \gamma $-rays, neutrons, and neutrinos, which arrive at the Earth, keep the information on the acceleration, because they are not affected by any magnetic field.
Therefore, neutral particles are useful to investigate the acceleration mechanism.
However, it is difficult to observe neutrinos, and it is difficult to distinguish whether $ \gamma $-rays have originated from ions or electrons.
Therefore, it is valuable to use neutrons in order to solve the acceleration mechanism of ions.
We note, however, that the survival probability of decay should be taken into account for neutrons of non-relativistic energy.

The Sun is the nearest source of particle acceleration to the Earth.
The Sun, which is a typical star, causes the explosive release of energy known as a solar flare.
The energy release in a flare amounts to $ 10^{29}-10^{33} {\rm \,ergs} $ during a few to a few tens of minutes.
The flare energy is released as electromagnetic radiation, plasma heating, and particle acceleration.
The information on ion acceleration is transferred to neutrons, line $ \gamma $-rays, and neutrinos which are produced by the interaction of accelerated particles with the solar atmosphere.

So far, $ \gamma $-rays have been observed in association with the solar flares by the Gamma Ray Spectrometer (GRS) onboard the Solar Maximum Mission (SMM) in the 1980s.
In the 1990s, solar $ \gamma $-rays were observed by detectors onboard the Compton Gamma Ray Observatory (CGRO) satellite, Oriented Scintillation Spectrometer Experiment (OSSE), Burst And Transient Source Experiment (BATSE), Energetic Gamma Ray Experiment Telescope (EGRET), and COMPton TELescope (COMPTEL).
Also since August 1991, the Gamma Ray Spectrometer (GRS) onboard the {\it Yohkoh} satellite has continued measurements.
The Reuven Ramaty High-Energy Solar Spectroscopic Imager (RHESSI) has started observations since February 2002. 

A lot of $ \gamma $-ray events associated with solar flares have been detected by these detectors \citep{Murphy1997}.
In some of these events, line $ \gamma $-rays, which originate from de-excited nuclei ($ 1 - 8 {\rm \,MeV} $), electron-positron annihilation ($ 0.511 {\rm \,MeV} $) and neutron capture ($ 2.223 {\rm \,MeV} $) processes, are observed with a continuous bremsstrahlung component.
The de-excited nuclei $ \gamma $-rays indicate the occurrence of ion acceleration, and $ 2.223 {\rm \,MeV} $ $ \gamma $-rays are produced by neutrons.
However line $ \gamma $-rays cannot be clearly observed in many events because they are buried in the bremsstrahlung component.

The importance of observing solar neutrons produced by accelerated particles was pointed out by \citet{Biermann1951} and by \citet{Lingenfelter1965a,Lingenfelter1965b}.
Lower energy neutrons (kinetic energy below $ 100 {\rm \,MeV} $) are observed only in space because they are attenuated in the Earth's atmosphere.
And high energy neutrons with kinetic energies higher than $ 100 {\rm \,MeV} $ are observed on the ground.
By simultaneous observations in space and on the ground, an energy spectrum of solar neutrons, or accelerated particles, can be obtained in a wide energy range.
Moreover, by acquiring the knowledge of the acceleration time, we can know if particle acceleration has taken place impulsively or gradually. 

Solar flares which produce neutrons occur frequently when solar activity is the maximum in a solar cycle, and they have been X-class solar flares in most cases.
Solar neutrons were observed for the first time by {\it SMM}/GRS at the flare on June 21, 1980 \citep{Chupp1982}.
On June 3, 1982, solar neutrons were detected for the first time by the ground-level neutron monitor installed at Jungfraujoch, Switzerland, as well as by {\it SMM}/GRS \citep{Chupp1987}.
In this event, high energy solar neutrons were observed by the IGY type neutron monitor, and low energy neutrons and high energy $ \gamma $-rays were observed by the {\it SMM}/GRS satellite at the same time.
Since a long tail of signals was observed by the neutron monitor, this event was interpreted using the impulsive and gradual particle acceleration models, and the neutron energy spectrum of this event was described as a power law with index $ = -2.4 $.
But, by \citet{Shibata1993b}, the time-extended neutron signals were interpreted using only the impulsive model.
This distinction came from the difference of the neutron propagation model in the Earth's atmosphere.

After the solar neutron event on June 3, 1982, four solar neutron events were observed with ground-based detectors.
The second event was observed by the IGY type neutron monitor located at Climax and a lot of stations in North America on May 24, 1990 \citep{Shea1991,Debrunner1997,MurakiShibata1996}.
The third event was on March 22, 1991 \citep{PyleSimpson1991}.
This event was observed by the NM64 type neutron monitor at Haleakala, Hawaii.
And the fourth and fifth events were on June 4 and 6, 1991 \citep{Muraki1992,Struminsky1994}.
On June 4, solar neutron signals were recorded by three different detectors (neutron monitor, neutron telescope, and muon telescope) located at Mt. Norikura.
On June 6, solar neutrons were detected at Haleakala and Mt. Norikura, at the same time.
Until the present time, solar neutrons have only been observed in association with solar flares larger than the X8 class \citep{Shibata1993a}.

In this paper, we report detection of solar neutrons by the neutron monitor installed at Mt.\,Chacaltaya, Bolivia.
Solar neutrons were produced in association with a solar flare which occurred at 14:51\,UT on November 24, 2000.
The soft X-ray class was X2.3.
Therefore, this event is not only the first report of the detection of solar neutrons by a neutron monitor in the 23rd solar cycle, but also the first ground-level event recorded with a soft X-ray class smaller than X8.
The analysis result of the neutron monitor data for this solar flare is described.
It is compared with X-ray and $ \gamma $-ray data obtained by the {\it Yohkoh} satellite.

\section{OBSERVATION OF SOLAR NEUTRONS}
\label{event20001124}

Three X-class solar flares occurred successively in NOAA region 9236 on November 24, 2000.
At 14:51\,UT, an X2.3 class flare was observed which was the largest among these three.
The location of the active region was $ {\rm N} 22 ^{\circ} {\rm W} 07 ^{\circ} $ at the time of the flare, and the flare was a disk flare. 

\begin{figure}[tbp]
   \plotone{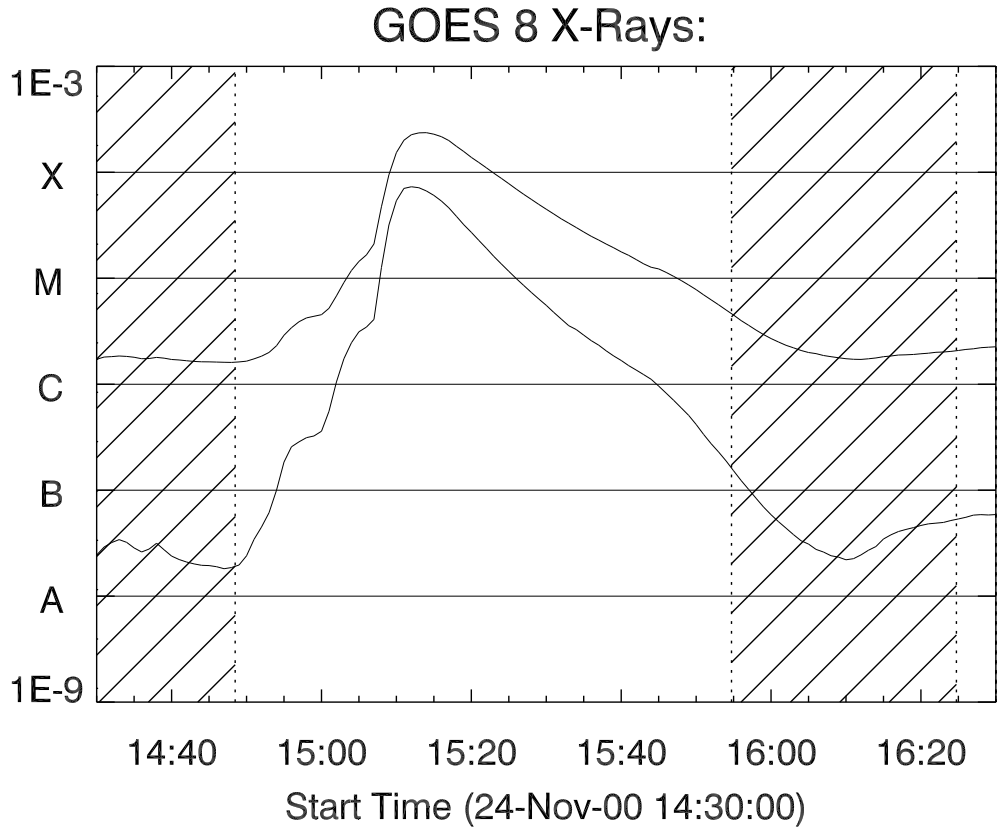}
   \caption{The time profile of the soft X-ray flux observed by the GOES satellite at the solar flare which occurred on November 24, 2000 (14:30$ - $16:30\,UT). The start time of this X2.3 flare was 14:51\,UT. Oblique lines mean night for the {\it Yohkoh} satellite. The upper profile expresses the X-ray flux in the wavelength range $ 1.0 - 8.0 {\rm \,\AA} $, and the lower one is for $ 0.5 - 4.0 {\rm \,\AA} $. \label{20001124_GOES}}
\end{figure}

The time profile of the soft X-ray flux in this flare observed by the GOES satellite is shown in Fig.\,\ref{20001124_GOES}.
Although the start time of this flare defined by the GOES satellite was 14:51\,UT, the time when the soft X-ray flux became a maximum was 15:13\,UT, which was more than 20 minutes after the start time.
In Fig.\,\ref{20001124_GOES}, we can distinguish three bumps starting at 14:51\,UT, 15:00\,UT, and 15:07\,UT, respectively.
Each of them arrives at C, M, and X class at its maximum.
In order to understand this stepwise increase of the soft X-ray flux, we examined the soft X-ray images obtained by the Soft X-ray Telescope (SXT) onboard the {\it Yohkoh} satellite.

\begin{figure}[tbp]
   \epsscale{0.35}
   \begin{center}
      \plotone{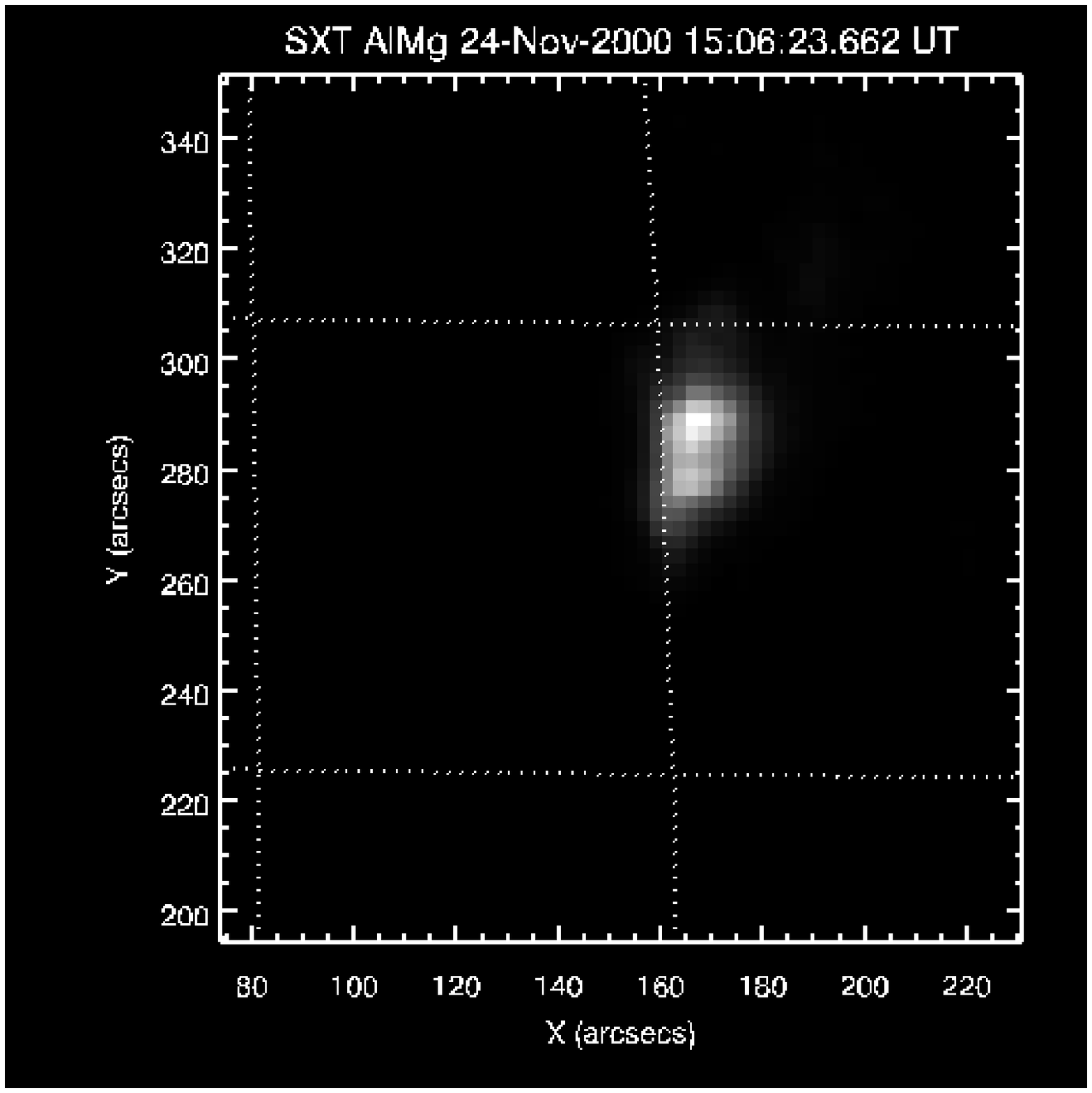}
   \end{center}
   \begin{center}
      \plotone{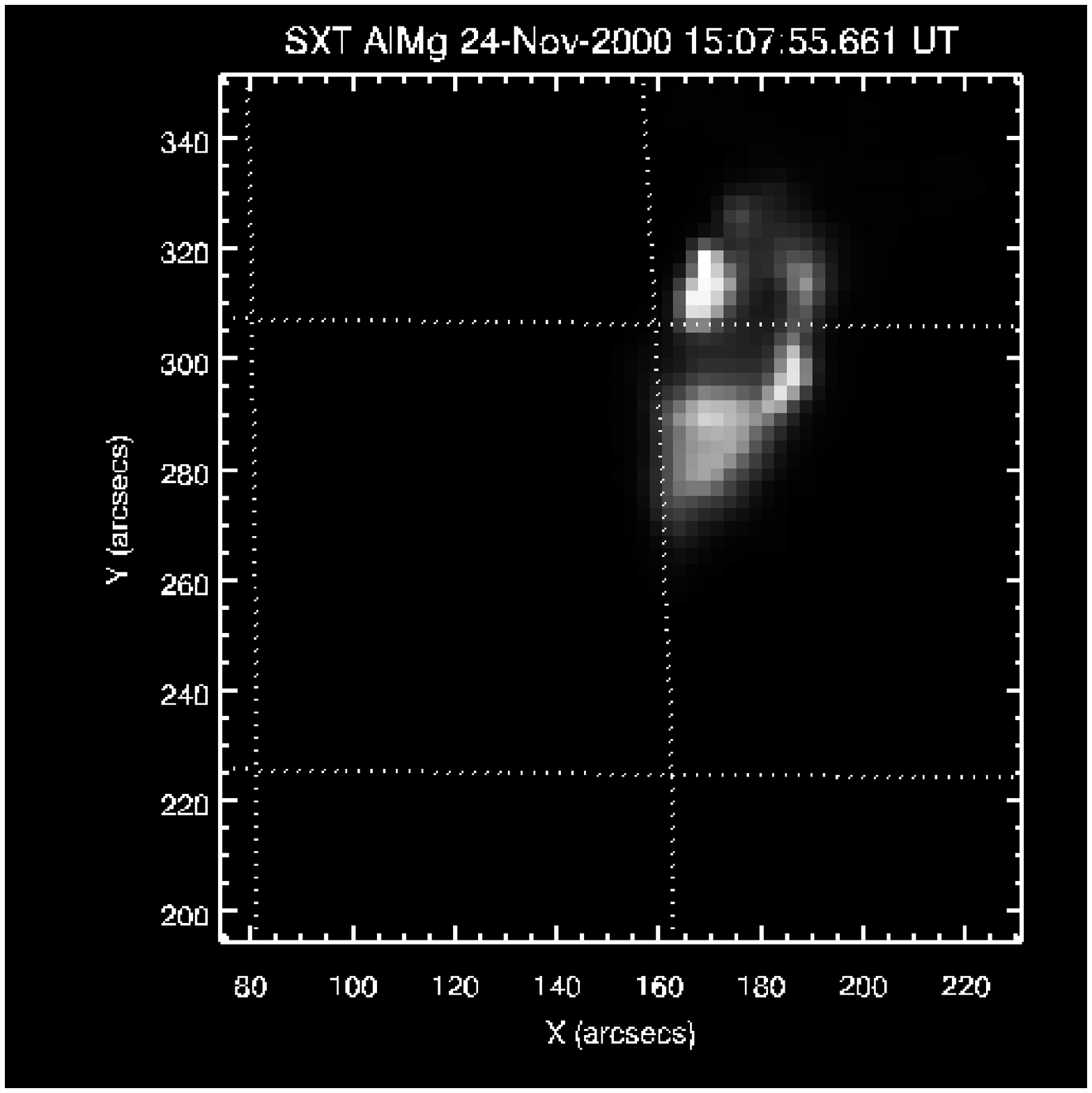}
   \end{center}
   \caption{Soft X-ray images observed by {\it Yohkoh}/SXT at 15:06:23\,UT and 15:07:55\,UT on November 24, 2000. The X-axis is in the $ {\rm E} - {\rm W} $ ($ {\rm E} $ is left) direction and the Y-axis is $ {\rm N} - {\rm S} $ ($ {\rm N} $ is top). At 15:06:23\,UT, the soft X-ray flux reached M class level, and at 15:07:55\,UT, it was X class level by the GOES satellite. In the bottom image, an isolated bright point appears which doesn't exist in the top image. \label{20001124_SXT}}
\end{figure}

Soft X-ray images observed by {\it Yohkoh}/SXT at 15:06:23\,UT and at 15:07:55\,UT on November 24, 2000 are shown in Fig.\,\ref{20001124_SXT}.
At 15:06:23\,UT, the soft X-ray flux reached M class level, and at 15:07:55\,UT, it was at X class level by the GOES satellite.
A bright point appears on the 15:07:55\,UT image, which didn't exist on the 15:06:23\,UT image.
From these images, we can say that distinct energy release phenomena occurred at around 15:07\,UT, and the soft X-ray flux observed by the GOES satellite was superimposed stepwise.

\begin{figure}[tbp]
   \begin{center}
      \includegraphics[scale=0.8]{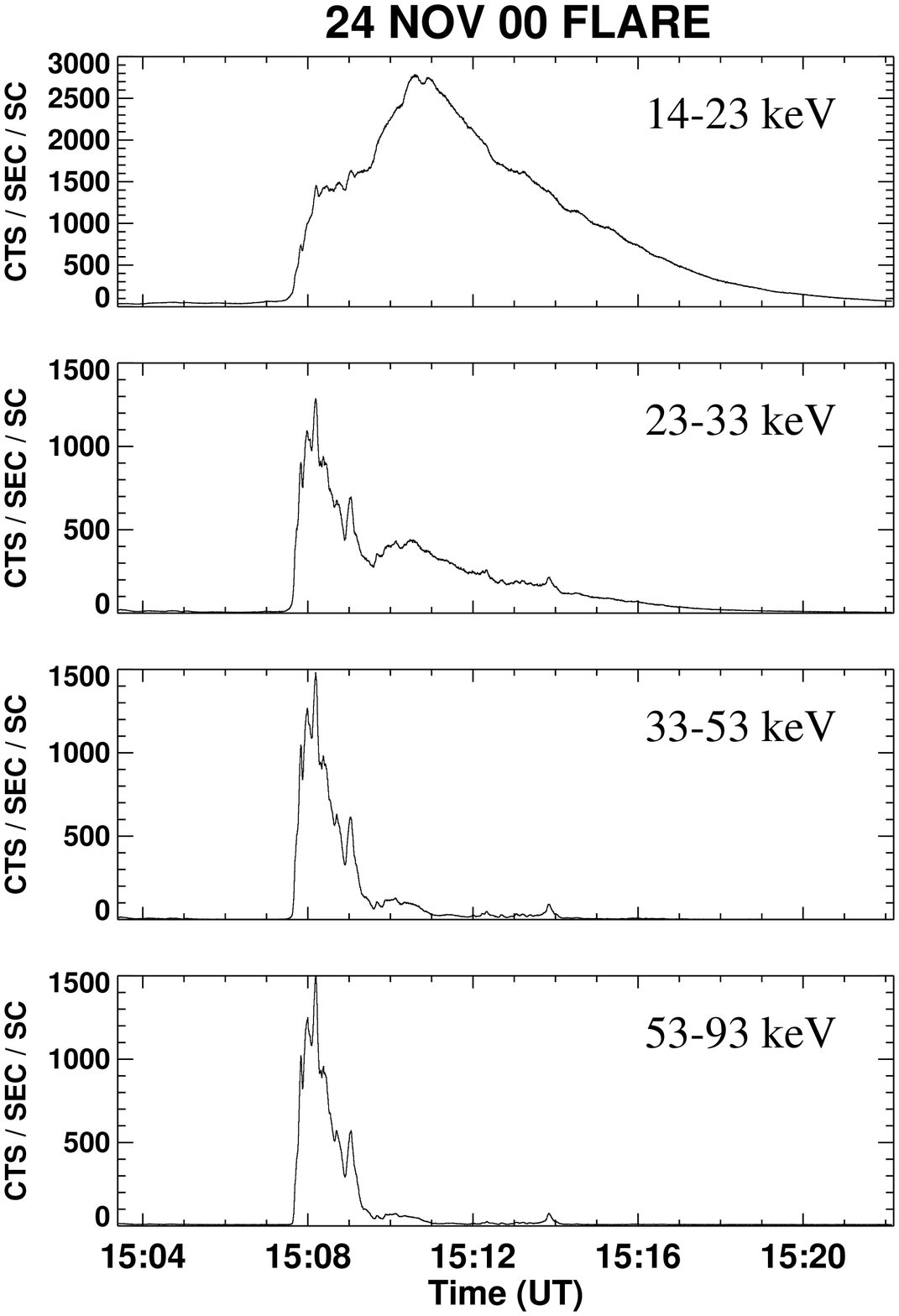}
   \end{center}
   \caption{Time profile of hard X-rays observed by {\it Yohkoh}/HXT on November 24, 2000. The vertical axis is the count rate per sub-collimator. Hard X-rays were suddenly emitted from 15:07:30\,UT. \label{20001124_HXT}}
\end{figure}

Around 15:08\,UT, when the energy release phenomenon occurred, a large amount of hard X-rays were observed.
The time profile of the hard X-rays observed by {\it Yohkoh}/HXT is shown in Fig.\,\ref{20001124_HXT}.
The hard X-ray flux suddenly increased from 15:07:30\,UT, when the new bright spot appeared in the soft X-ray image in Fig.\,\ref{20001124_SXT}.

\begin{figure}[tbp]
   \epsscale{0.35}
   \begin{center}
      \plotone{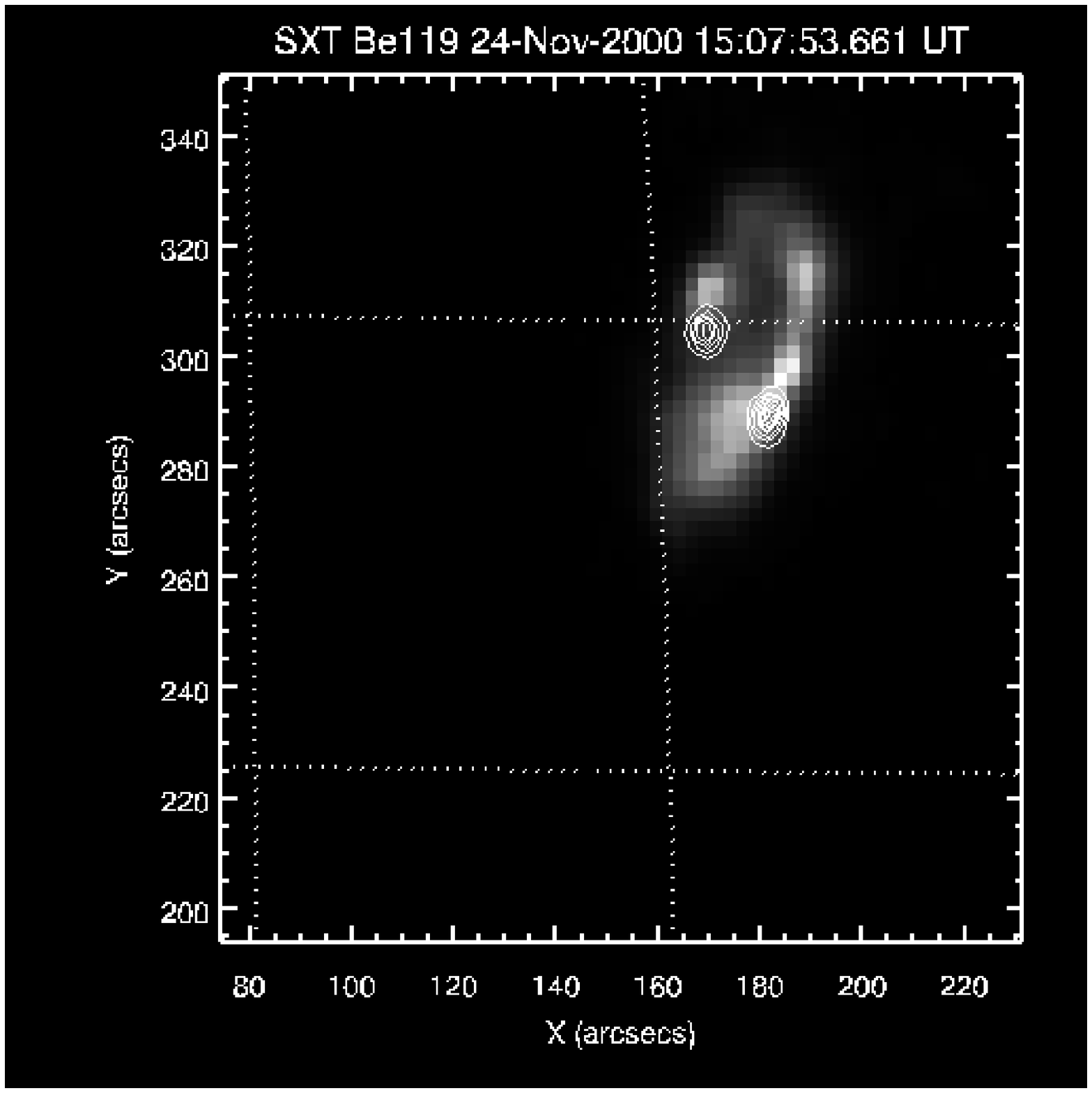}
   \end{center}
   \caption{Soft X-ray and hard X-ray images observed by {\it Yohkoh}/SXT and {\it Yohkoh}/HXT at 15:07:53\,UT on November 24, 2000. The hard X-ray image is drawn by contours on the soft X-ray image. The hard X-ray source exists at the foot point of the soft X-ray flare loop. \label{20001124_SXT_HXT}}
\end{figure}

The hard X-ray image observed by {\it Yohkoh}/HXT is shown in Fig.\,\ref{20001124_SXT_HXT} on the soft X-ray image observed by {\it Yohkoh}/SXT.
At the time when the new soft X-ray source appeared, the hard X-ray source turned up at the foot point of the soft X-ray flare loop.
Therefore, hard X-rays were produced in parallel with soft X-rays around 15:08\,UT.
Moreover, the energy spectrum of the hard X-rays was very hard between 15:07:30\,UT and 15:09:30\,UT.

High energy electrons can emit hard X-rays by bremsstrahlung.
Therefore, the onset time of hard X-rays can indicate the time when the particle acceleration took place.
But, hard X-rays are produced by high energy electrons, not by ions.
Therefore, we cannot assert that at the time when a large amount of hard X-rays are produced, ions are accelerated and solar neutrons are produced.

\begin{figure}[tbp]
   \epsscale{0.75}
   \plotone{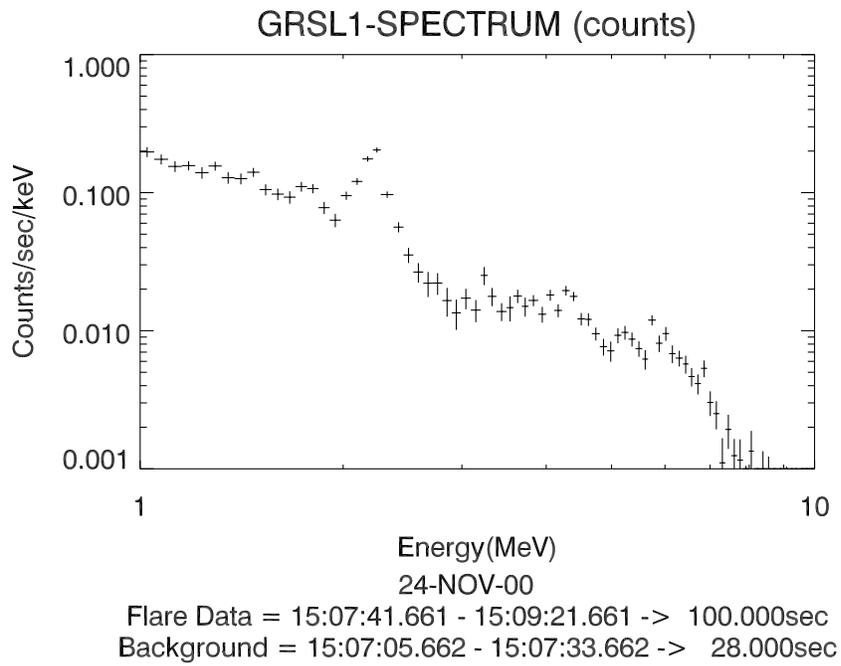}
   \caption{The background-subtracted $ \gamma $-ray spectrum observed by the {\it Yohkoh}/GRS on November 24, 2000. There is a clear signal of $ 2.223 {\rm \,MeV} $ $ \gamma $-rays and weak signals of de-excited nuclei line $ \gamma $-rays between $ 4 {\rm \,MeV} $ and $ 7 {\rm \,MeV} $ on the bremsstrahlung component. \label{20001124_GRS_spectrum}}
\end{figure}

It is more useful to examine information on $ \gamma $-rays to estimate the acceleration time of ions.
Around 15:08\,UT, a large amount of $ \gamma $-rays were observed by {\it Yohkoh}/GRS.
Fig.\,\ref{20001124_GRS_spectrum} shows the energy spectrum of $ \gamma $-rays at that time.
In this figure, a clear signal of $ 2.223 {\rm \,MeV} $ neutron capture line $ \gamma $-rays is seen on the bremsstrahlung component.
Therefore, it is evident that neutrons were produced around 15:08\,UT on November 24, 2000.
Furthermore, between $ 4 {\rm \,MeV} $ and $ 7 {\rm \,MeV} $, weak signals of line $ \gamma $-rays produced by de-excited ions, C($ 4.443 {\rm \,MeV} $) and O($ 6.129 {\rm \,MeV} $) for example, are seen too.
Consequently, it is certain that ions were accelerated at that time.

In principle, from these line $ \gamma $-ray data, we can calculate the amount of produced neutrons and the energy spectra of the accelerated ions independently of the neutron measurements \citep{Ramaty1996}.
But, at this event, the flux of line $ \gamma $-rays produced by de-excited ions was weak.
It is difficult to estimate the amount of $ \gamma $-rays produced by each de-excitation process due to a contamination of electron bremsstrahlung.
For this reason, the neutron spectrum was not derived from the line $ \gamma $-ray data to compare it with that determined by the neutron monitor data.

\begin{figure}[tbp]
   \epsscale{0.75}
   \plotone{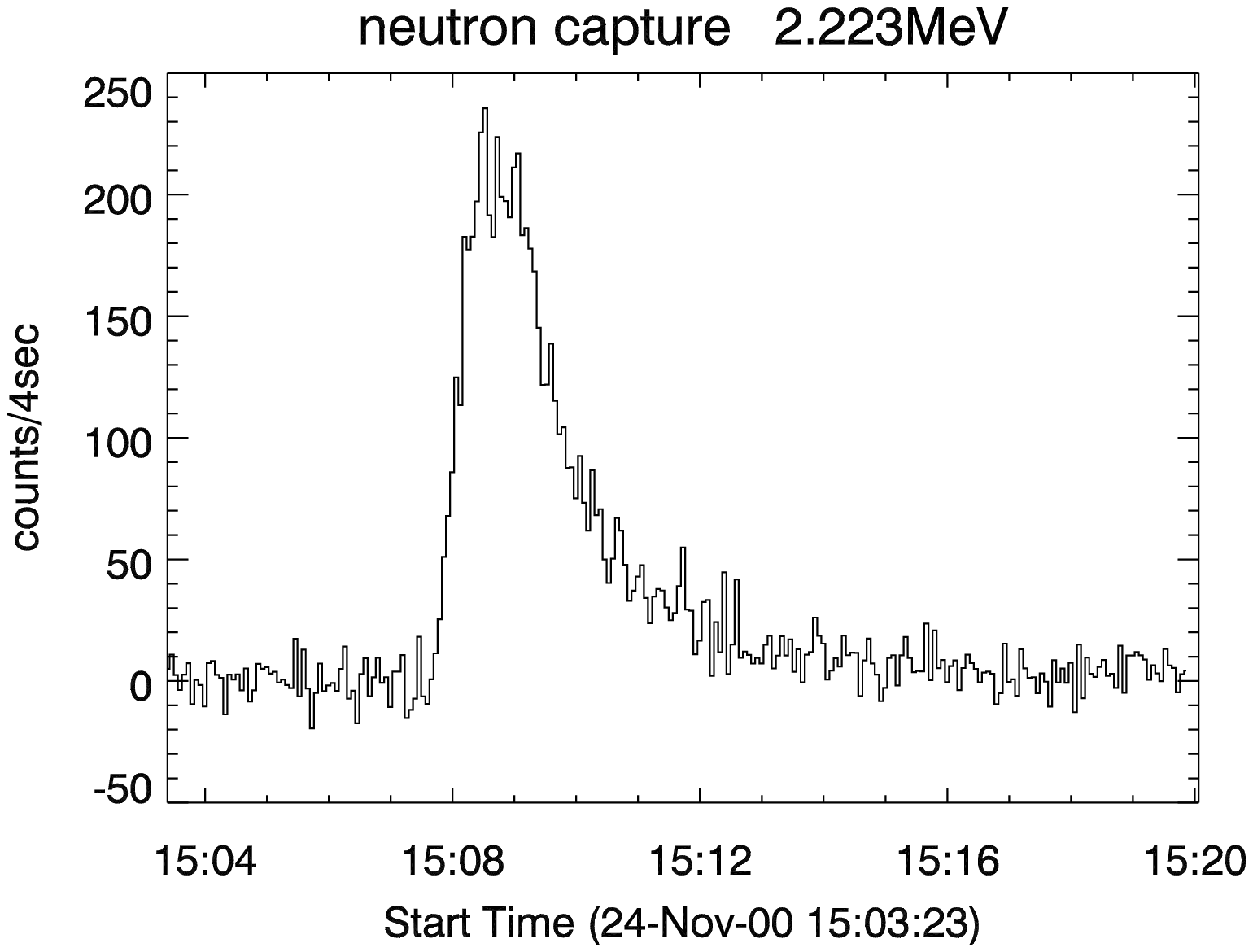}
   \plotone{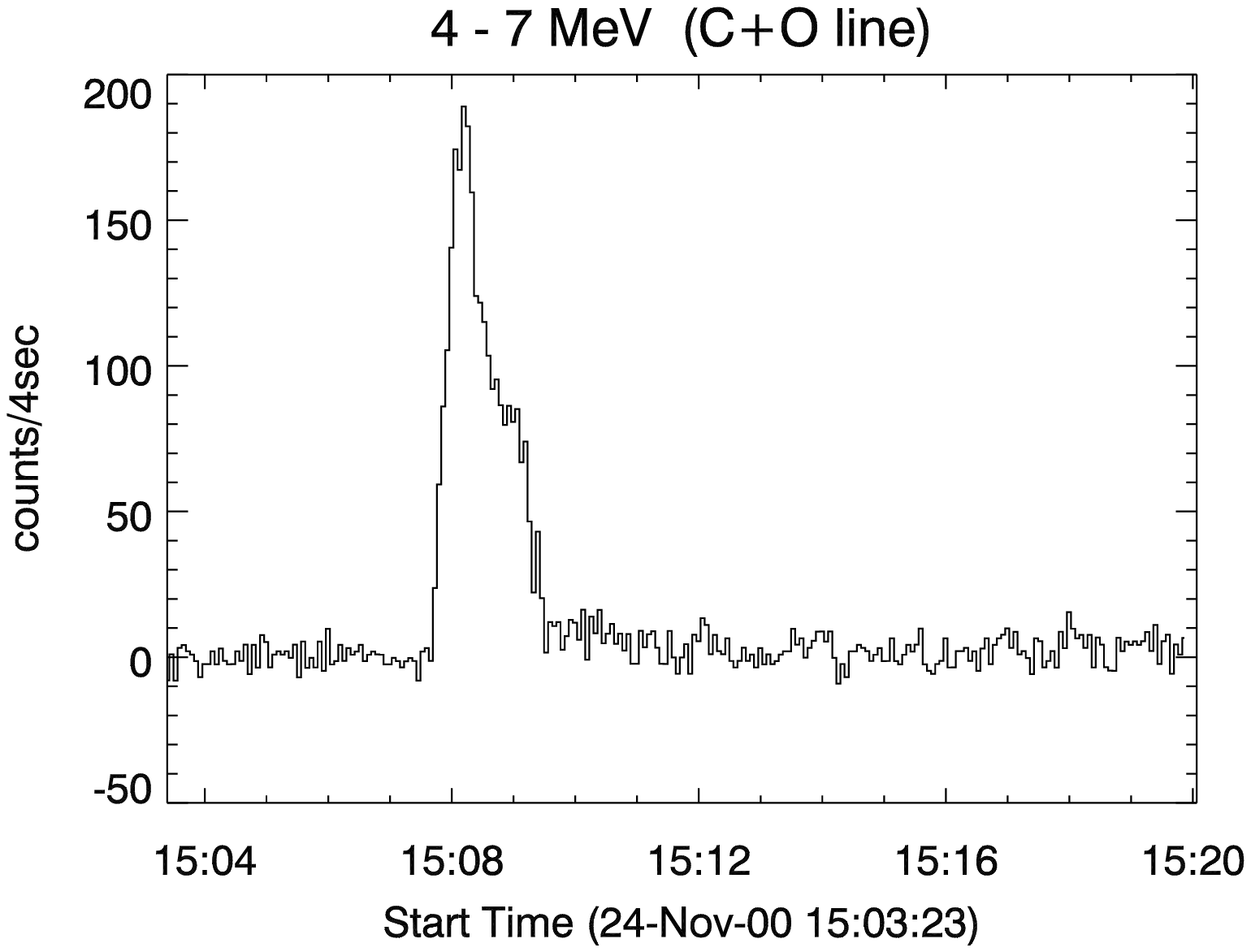}
   \caption{The time profiles of $ 2.223 {\rm \,MeV} $ neutron capture line $ \gamma $-rays and $ 4-7 {\rm \,MeV} $ $ \gamma $-rays on November 24, 2000. The upper figure is the $ 2.223 {\rm \,MeV} $ time profile, and the lower one is the $ 4-7 {\rm \,MeV} $ time profile. The bremsstrahlung component is not subtracted. The $ 4-7 {\rm \,MeV} $ $ \gamma $-ray time profile includes line $ \gamma $-rays produced by de-excited ions, C($ 4.443 {\rm \,MeV} $) and O($ 6.129 {\rm \,MeV} $). In both of them, the vertical axis is the counting rate per 4 seconds. The time profile of neutron capture $ \gamma $-rays is more expanded than that of $ 4-7 {\rm \,MeV} $ $ \gamma $-rays. \label{20001124_GRS_time_prof}}
\end{figure}

The time profiles of $ 2.223 {\rm \,MeV} $ neutron capture line $ \gamma $-rays and $ 4-7 {\rm \,MeV} $ $ \gamma $-rays observed by the {\it Yohkoh}/GRS around 15:08\,UT on November 24, 2000, are shown in Fig.\,\ref{20001124_GRS_time_prof}.
The bremsstrahlung component is not subtracted.
The $ 4-7 {\rm \,MeV} $ $ \gamma $-ray time profile included line $ \gamma $-rays produced by de-excited ions, C($ 4.443 {\rm \,MeV} $) and O($ 6.129 {\rm \,MeV} $).
Therefore, the $ 4-7 {\rm \,MeV} $ $ \gamma $-ray time profile is approximately equal to the C+O line and bremsstrahlung $ \gamma $-rays time profile.
In these figures, the duration of $ 2.223 {\rm \,MeV} $ $ \gamma $-ray emission is longer than that of $ 4-7 {\rm \,MeV} $ $ \gamma $-rays.
The $ 2.223 {\rm \,MeV} $ $ \gamma $-rays are produced when thermal neutrons are captured by hydrogen.
High energy neutrons are produced simultaneously with line $ \gamma $-rays of de-excited ions by interactions of accelerated ions and the solar atmosphere.
On the other hand, the $ 2.223 {\rm \,MeV} $ $ \gamma $-rays are produced about $ 100 {\rm \,s} $ after the production of the high energy neutrons, because of the time required for neutrons to slow down and be captured.
Therefore, the time profile of $ 2.223 {\rm \,MeV} $ $ \gamma $-rays is dilated compared with that of $ 4-7 {\rm \,MeV} $ $ \gamma $-rays.
Hence, we can consider that the high energy neutrons were produced at the same time as de-excited nuclei $ \gamma $-rays rather than $ 2.223 {\rm \,MeV} $ $ \gamma $-rays.

At 15:08\,UT on November 24, 2000, the Sun was over Bolivia.
The neutron monitor installed at Mt.\,Chacaltaya, Bolivia was at the most suitable place for observing solar neutrons.
This station is located at $ 292.0 ^{\circ} {\rm E}$, $ 16.2 ^{\circ} {\rm S} $, $ 5250 {\rm \,m} $ above sea level, and the vertical air mass is $ 540 {\rm \,g/cm} ^2 $. 
At this time, the zenith angle of the Sun was $ 17.47 ^{\circ} $ and the air mass for the line of sight to the Sun was $ 566 {\rm \,g/cm} ^2 $.

\begin{figure}[tbp]
   \epsscale{1.0}
   \plotone{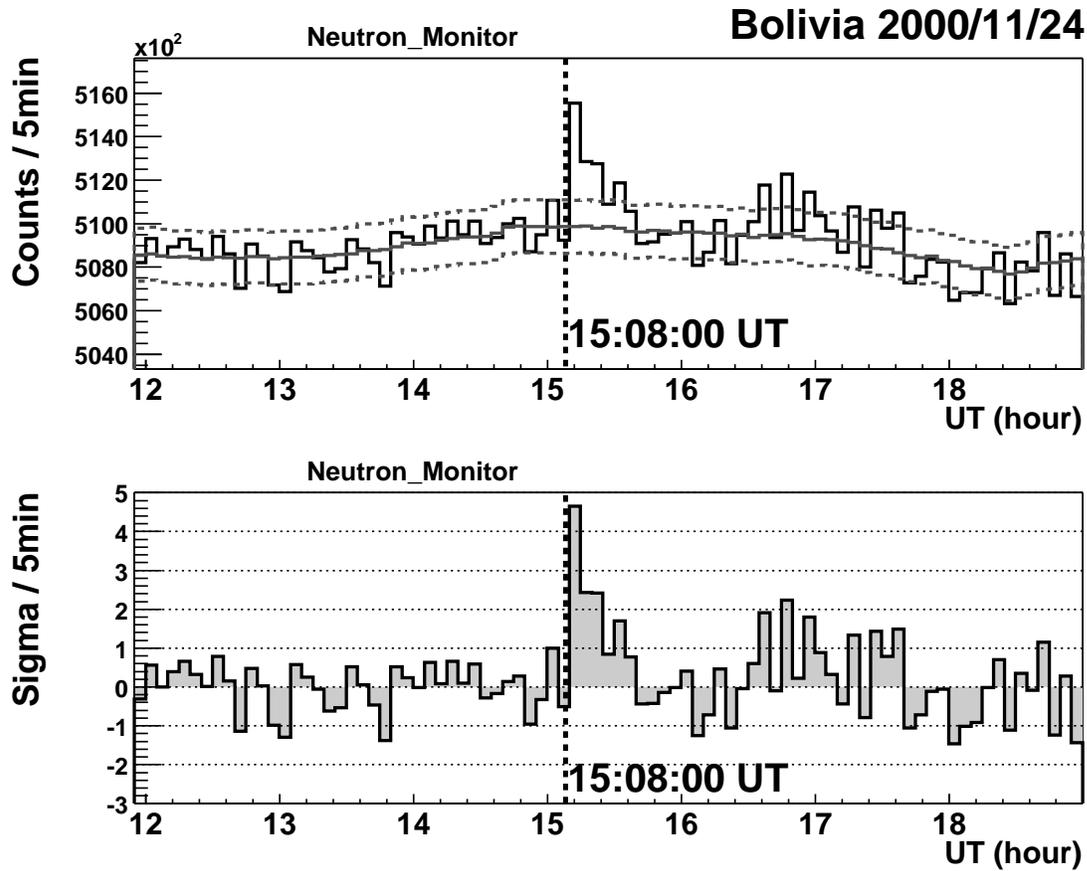}
   \caption{The 5 minute counting rate observed by the Bolivia neutron monitor on November 24, 2000. This is the pressure-corrected data. The top panel is the time profile observed by the neutron monitor and background. The solid smooth line is the averaged background, and dashed lines are $ \pm 1 \,\sigma $ from the background. The bottom panel is the statistical significance. From 15:10\,UT to 15:25\,UT, a clear excess which was made by solar neutrons was observed. The total statistical significance for 15 minutes is $ 5.5 \,\sigma $. \label{20001124_NM}}
\end{figure}

The neutron monitor installed at Mt.\,Chacaltaya is $ 13.1 {\rm \,m}^2 $ in area and of the NM64 type.
The counting rate is recorded every 10 seconds.
The time profile of neutrons observed by the neutron monitor is shown in Fig.\,\ref{20001124_NM}.
A clear excess was found between 15:10\,UT and 15:25\,UT.
The statistical significance of each bin is $ 4.7 \,\sigma $ at 15:10$ - $15:15\,UT, $ 2.4 \,\sigma $ at 15:15$ - $15:20\,UT and $ 2.4 \,\sigma $ at 15:20$ - $15:25\,UT.
The total significance for 15 minutes, between 15:10\,UT and 15:25\,UT, turned out to be $ 5.5 \,\sigma $.

There is a possibility that these excesses came from energetic ions because the neutron monitor can observe energetic ions.
But, there is no evidence that the enhancement was produced by energetic ions since the measurements by the other stations in the worldwide network of neutron monitors showed no enhancement.
In addition, the cutoff rigidity at Mt.\,Chacaltaya is high, that is $ 12.53 {\rm \,GV}$, so it is difficult for ions to reach ground level.
There are many neutron monitor stations located at the place where are lower cutoff rigidity.
If these excesses came from energetic ions, some enhancement should have been detected by other stations.
And protons with kinetic energy greater than $ 100 {\rm \,MeV} $ were observed by the GOES satellite in association with the X2.3 solar flare, which was observed 1 hour later.
Therefore, these signals must have come from solar neutrons.

Neutron monitors cannot measure the energy of neutrons.
Therefore, we cannot directly derive the energy spectrum of solar neutrons.
But, using the time of flight (TOF) method by assuming the emission time of solar neutrons, this can be derived. 
To this end, we must postulate the production time of solar neutrons.
As shown above, the production time of de-excited nuclei line $ \gamma $-rays is taken to be that of solar neutrons.
Unfortunately, the excess obtained in this event is not strong enough to examine the time profile of the signals in detail.
Therefore we do not touch on the possibility of an extended production of neutrons in this paper and simply assume that neutrons were produced instantaneously in this event.

If solar neutrons were produced at 14:51\,UT, which was the start time of soft X-rays, then the energy of the neutrons detected between 15:10\,UT and 15:25\,UT is calculated to be $ 47-19 {\rm \,MeV} $.
Since neutrons suffer violent attenuation in the Earth's atmosphere, such low energy neutrons cannot arrive at the ground.
On the other hand, the assumption of the production time as 15:08\,UT, which is the time hard X-ray and $ \gamma $-ray emissions were seen, gives $ 772-57 {\rm \,MeV} $, which is much more reasonable.
Consequently, the production time of solar neutrons is set at 15:08\,UT hereinafter.

From the time profile obtained by the neutron monitor, the flux of solar neutrons at the top of the Earth's atmosphere was calculated by the following formula:
\begin{equation}
\frac{ \Delta N }{ \epsilon \times S \times \Delta E_n } \label{eq:flux}
\end{equation}
where $ \Delta N $ is the excess count contributed by solar neutrons, and $ \epsilon $ is the detection efficiency of the neutron monitor.
Here, $ \epsilon $ includes the attenuation of solar neutrons through the Earth's atmosphere.
$ S $ is the area of the neutron monitor, and $ \Delta E_n $ is the energy range corresponding to one time bin.

\begin{figure}[tbp]
   \begin{center}
      \includegraphics[angle=-90,scale=0.6]{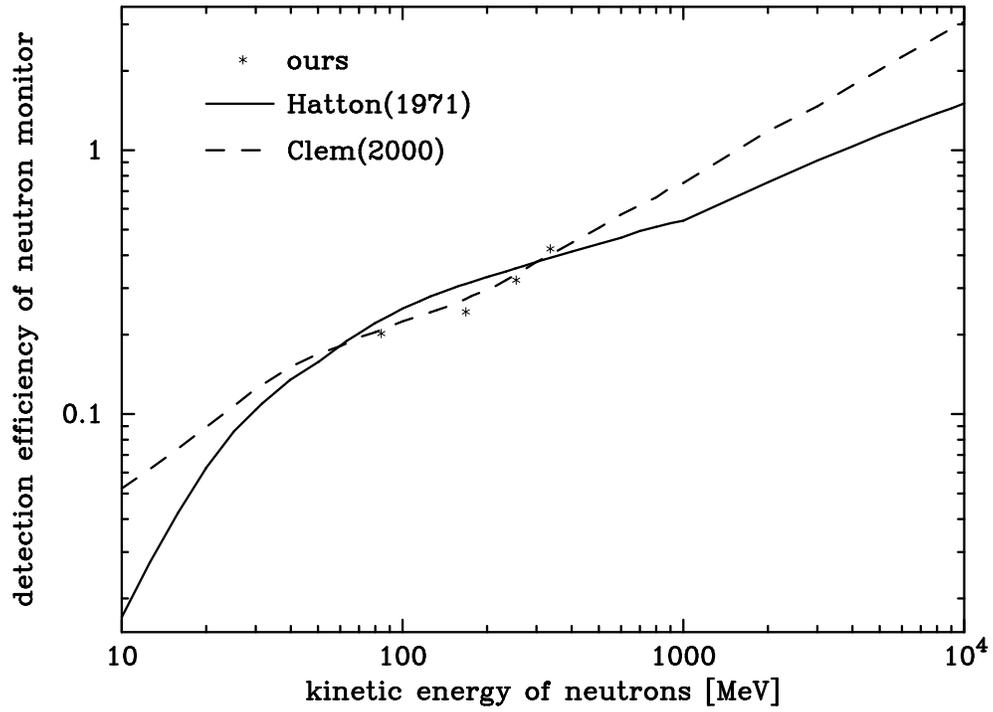}
   \end{center}
   \caption{The detection efficiency of the neutron monitor \citep{Shibata2001}. In this diagram `ours' is the result of the accelerator experiments by \citet{Shibata2001}. The solid line is the Monte Carlo prediction of \citet{ClemDorman2000}. The dashed line is the value of \citet{Hatton1971}.  \label{NM_Efficiency}}
\end{figure}

The detection efficiency of the neutron monitor and the attenuation of solar neutrons depend on the energy of the neutrons.
The detection efficiency of the neutron monitor was calculated by \citet{Hatton1971}, and recently by \citet{ClemDorman2000}.
Experimentally, that has been measured directly by the accelerator experiment \citep{Shibata2001}.
The results are shown in Fig.\,\ref{NM_Efficiency} together with calculations by \citet{Hatton1971}, and by \citet{ClemDorman2000}.
Although the two calculations are consistent with the experimental result in the energy range $ 100 - 400 {\rm \,MeV} $, there is a big discrepancy between the calculation of \citet{Hatton1971} and \citet{ClemDorman2000} outside this experimental range.
The deviation of the calculation by \citet{ClemDorman2000} from the experimental result is $ \pm 5 \,\% $, whereas that by \citet{Hatton1971} is $ \pm 10 \,\% $.
Consequently, in Equation (\ref{eq:flux}), we adopted the efficiency calculated by \citet{ClemDorman2000} because it was closer to the result of \citet{Shibata2001} than the calculation by \citet{Hatton1971}.

The attenuation of solar neutrons in the Earth's atmosphere was calculated by \citet{Debrunner1989} and \citet{Shibata1994} by Monte Carlo simulations.
Hereinafter they are called the Debrunner model and the Shibata model, respectively.
There is a big discrepancy between the two models.
In order to examine which model is correct, the accelerator experiment was conducted at the Research Center for Nuclear Physics, Osaka University (RCNP).
The Shibata model can explain the experimental result \citep{Koi2001}.
Consequently, we adopted the Shibata model in calculating the propagation of neutrons in the air.

\begin{figure}[tbp]
   \begin{center}
      \includegraphics[angle=-90,scale=0.6]{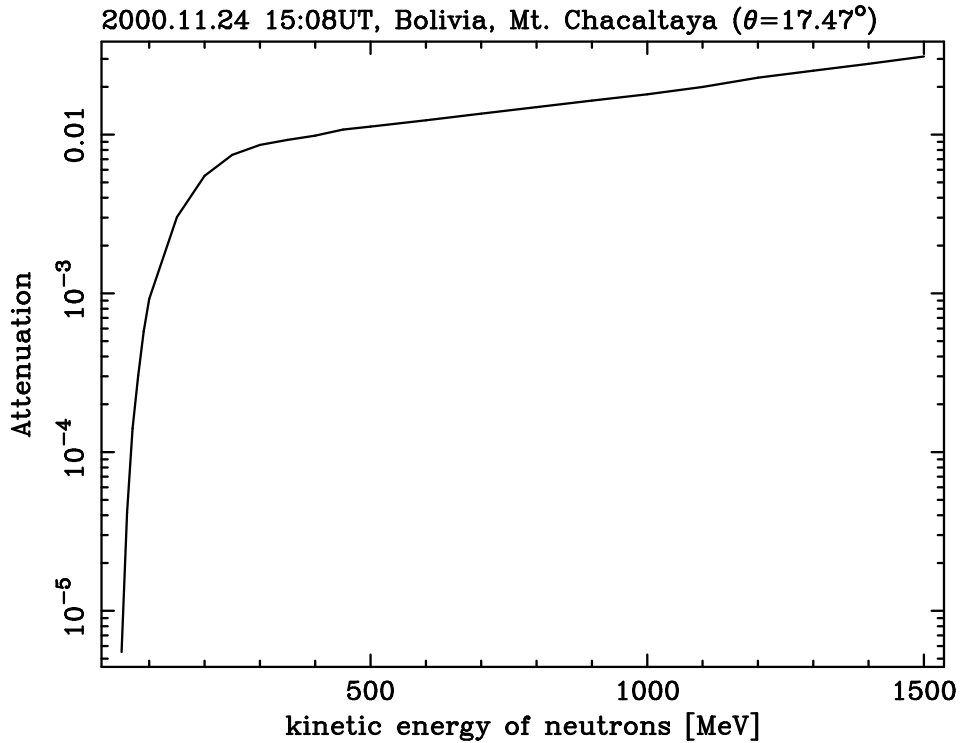}
   \end{center}
   \caption{The attenuation of solar neutrons through the Earth's atmosphere at Mt.\,Chacaltaya at 15:08\,UT on November 24, 2000. This was obtained by using the Shibata model. The horizontal axis is the kinetic energy of the neutrons at the top of the Earth's atmosphere.\label{20001124_Atte}}
\end{figure}

The attenuation of solar neutrons in the Earth's atmosphere at the ground level of Mt.\,Chacaltaya at 15:08\,UT on November 24, 2000 is shown in Fig.\,\ref{20001124_Atte}.
This is obtained by using the Shibata model.
Solar neutrons whose kinetic energy are below $ 100 {\rm \,MeV} $ are strongly attenuated by the Earth's atmosphere.

\begin{figure}[tbp]
   \begin{center}
      \includegraphics[scale=0.6]{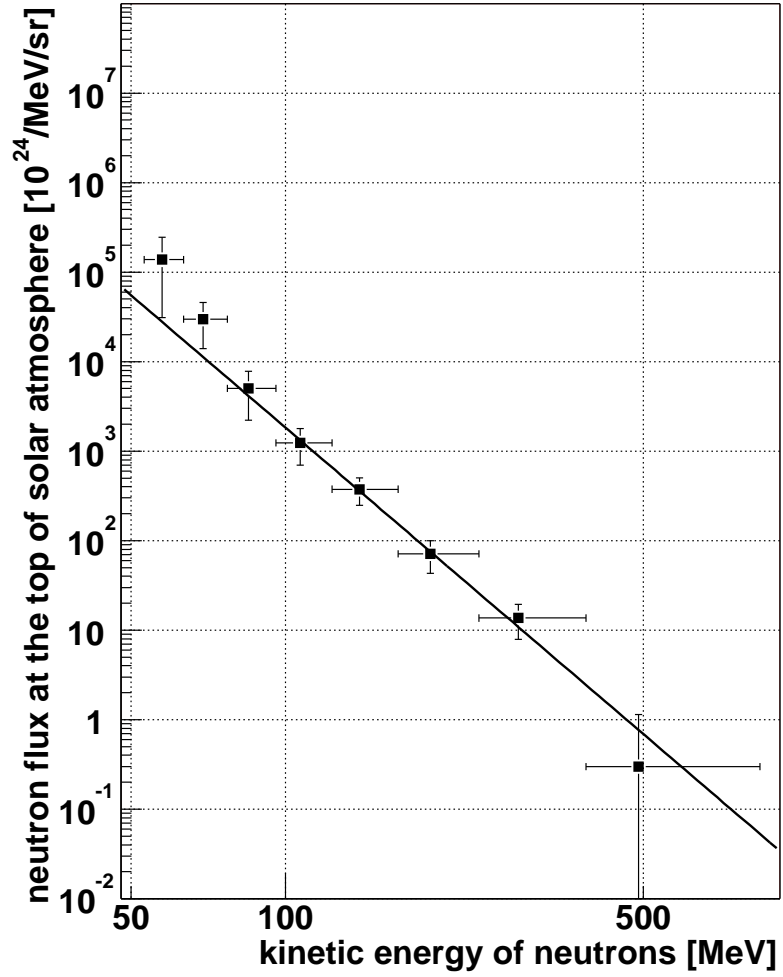}
   \end{center}
   \caption{The energy spectrum of solar neutrons on the solar surface for the flare that occurred on November 24, 2000. \label{20001124_Flux}}
\end{figure}

To derive the energy spectrum of neutrons at the solar surface from the flux at the top of the Earth's atmosphere, the survival probability of neutrons between the Sun and the Earth is taken into account.
The result is shown in Fig.\,\ref{20001124_Flux}.
The vertical errors come from only statistical ones.
This spectrum was derived from 2 minutes counting rate.
By fitting these data points with a power law of the form $ C \times (E_n/100[{\rm MeV}])^{\alpha} $, the energy spectrum of solar neutrons was obtained.
$ C $ is the flux of neutrons at $ 100 {\rm \,MeV} $, and $ \alpha $ means the power law index.
The fitting region is chosen above $ 100 {\rm \,MeV} $, because there the errors from neutron attenuation in the Earth's atmosphere are small.
These values were obtained as follows:
\begin{eqnarray}
(1.8 \pm 0.8) \times 10^{27} \, {\rm [/MeV/sr]} \times \left( \frac{E_n}{100 \, {\rm [MeV]}} \right)^{-4.9 \, \pm \, 0.7}.
\label{eq:flux_20001124}
\end{eqnarray}
For this fit, the value of $ \chi ^2 $/dof = 0.307/3.
The total energy flux of solar neutrons which were emitted from the Sun with energy range between $ 50 - 800 {\rm \,MeV} $ was calculated as $ 7.4 \times 10^{25} {\rm \,erg/sr} $.
This is obtained by simply integrating Equation (\ref{eq:flux_20001124}).
We did not assume any turnover of the energy spectrum of neutrons in this energy region.

\section{DISCUSSION}
\label{discussion}

\begin{table}
\footnotesize
   \caption[Solar neutron events]{Solar neutron events}
      \begin{tabular}{c||c|c|c|c|c} \hline
Date & June 3 1982 & May 24 1990 & March 22 1991 & June 4 1991 & November 24 2000 \\ \hline
Time [UT] & 11:43 & 20:48 & 22:44 & 3:37 & 14:51 \\ \hline
Observatory & Jungfraujoch & Climax & Haleakala & Mt. Norikura & Mt.\,Chacaltaya \\
& (Switzerland) & (USA) & (Hawaii, USA) & (Japan) & (Bolivia) \\
Height [m] & $ 3475 $ & $ 3400 $ & $ 3030 $ & $ 2770 $ & $ 5250 $ \\ \hline
X-ray class & X8.0 & X9.3 & X9.4 & X12.0 & X2.3 \\ \hline
Sunspot location & S09 E72 & N36 W76 & S26 E28 & N30 E70 & N22 W07 \\ \hline
Detector & 12IGY & 12IGY & 18NM64 & 12NM64 & 12NM64 \\ \hline
Flux at $ 100 {\rm \,MeV} $ &  &  &  &  & \\
$ [\times 10^{28} {\rm \,neutrons} $ & $ 2.6 \pm 0.7 $ & $ 4.3 \pm 0.4 $ & $ 0.06 \pm 0.01 $ & $ 1.8 \pm 0.2 $ & $ 0.18 \pm 0.08 $ \\
\hspace*{10mm} $ {\rm /MeV/sr}] $ &  &  &  &  & \\ \hline
Power index & $ -4.0 \pm 0.2 $ & $ -2.9 \pm 0.1 $ & $ -2.7 \pm 0.1 $ & $ -7.3 \pm 0.2 $ & $ -4.9 \pm 0.7 $ \\ \hline
    \end{tabular}
  \label{neutron_event}
\end{table}%

Energy spectra of solar neutron events at the top of Earth's atmosphere were provided by \citet{Shibata1993a}.
They used the attenuation of solar neutrons in the atmosphere calculated by \citet{Shibata1993b} and the neutron monitor detection efficiency calculated by \citet{Hatton1971}.
Using the \citet{Shibata1993a} values, neutron energy spectra at the solar surface of these solar neutron events can be calculated.
Calculated values are shown in Table \ref{neutron_event} together with the results of the November 24, 2000 event discussed in this paper.
The solar neutron event on June 4, 1991 was observed by three different detectors, but only the neutron monitor value is shown in Table \ref{neutron_event}.
The data for the  solar neutron event on June 6, 1991, do not exist in \citet{Shibata1993a}.
The solar neutron event on March 22, 1991, is the weakest event in Table \ref{neutron_event}.
Because the Haleakala neutron monitor has a large area, this event was detected.
Comparing the solar neutron events observed in former cycles with that observed on November 24, 2000, the latter event is fainter than previous events.
Because of the thin air mass at Mt.\,Chacaltaya ($ 540 {\rm \,g/cm^2} $ for vertical), even faint signals can be detected.

Table \ref{neutron_event} shows most solar neutron events observed in former solar cycles come from limb flares.
But, the November 24, 2000 event came from disk flare.
Solar neutrons are released tangentially to the solar surface \citep{HuaLingenfelter1987a,HuaLingenfelter1987b,Hua2002}.
Therefore, solar neutrons are thought to be detected easily from limb flares.
The result in the solar neutron event in solar cycle 23, however, is not explained by this argument.
The production direction of solar neutrons in the solar atmosphere has to be treated in more detail taking into account the position of the solar flare at the solar surface and the loop structure of each event.

The spectrum of accelerated ions can be calculated from the neutron spectrum using the spectrum of escaping neutrons produced by the accelerated ions \citep{HuaLingenfelter1987a,HuaLingenfelter1987b}.
From the neutron spectra shown in Table \ref{neutron_event}, the number of protons above $ 30 {\rm \,MeV} $ would be larger than $ 10 ^{33} $ in all events under the assumption that there is no turnover of the spectrum.
This is larger than the number of protons accelerated in the very large flare on June 4, 1991 \citep{Murphy1997}.
Estimating the number of solar neutrons below $ 100 {\rm \,MeV} $ accurately, however, is impossible because these low energy neutrons suffer from high attenuation through the Earth's atmosphere.
Possibly the turnover of the spectrum in the low energy range results in a smaller estimate of the number of accelerated protons.
The observation of solar neutrons below $ 100 {\rm \,MeV} $ by satellite experiments is indispensable in order to determine the total number of protons accelerated in a solar flare.

\section{CONCLUSIONS}
\label{conclusion}

We detected solar neutrons in association with the solar flare occurred on November 24, 2000.
This detection was made by the neutron monitor at Mt.\,Chacaltaya, where was a very suitable site to observe solar neutrons in this flare.

Assuming neutrons were produced impulsively at 15:08\,UT, the energy spectrum of solar neutrons at the solar surface was obtained as
\begin{eqnarray*}
(1.8 \pm 0.8) \times 10^{27} \, {\rm [/MeV/sr]} \times \left( \frac{E_n}{100 \, {\rm [MeV]}} \right)^{-4.9 \, \pm \, 0.7}
\end{eqnarray*}
and the total integrated energy of solar neutrons in the energy range $ 50 - 800 {\rm \,MeV} $ emitted from the Sun was calculated to be $ 7.4 \times 10^{25} {\rm \,erg/sr} $.

We assumed neutrons were produced at 15:08\,UT because large amount of hard X-rays and $ \gamma $-rays were observed.
If neutrons were produced at the start time of the solar flare defined by the GOES satellite (14:51\,UT), the energy of neutrons was estimated to be too low to be detected on the ground due to the extreme attenuation in the Earth's atmosphere.
Therefore, it can be said that neutrons are not necessarily produced following the time profile of soft X-rays.

In order to investigate the production time of neutrons, we compared the solar neutron data with the X-ray and $ \gamma $-ray data obtained by the {\it Yohkoh} satellite.
From the data of {\it Yohkoh}/HXT and {\it Yohkoh}/GRS, hard X-rays and $ \gamma $-rays were observed with strong intensity.
The observability of solar neutrons is possibly correlated with the intensity of hard X-rays and $ \gamma $-rays in a solar flare.

Most solar neutron events observed before the 23rd solar cycle were limb flares, and theoretically solar neutrons were thought to be detectable only from limb flares.
But, the flare discussed in this paper was a disk flare, and by combining our result with former events, the detectability of solar neutrons seems to be independent of the position of the flare on the Sun.
The number of solar neutron events is, however, still too small.
For the statistical argument, it is necessary to increase the number of solar neutron events.
At the same time, further study of the acceleration of particles and the production and propagation of neutrons at the solar surface for each event is necessary, for example, by Monte Carlo simulation, including real loop construction of each solar flare.

All the solar neutron events observed in the 21st and 22nd solar cycles were associated with solar flares beyond X8.
However, the X-ray class was X2.3 on November 24, 2000, when solar neutrons were observed.
This is the first report of the detection of solar neutrons on the ground associated with a solar flare whose X-ray class is smaller than X8.

\acknowledgments

The authors wish to thank the {\it Yohkoh} team, for their support to the mission and guidance in the analysis of the {\it Yohkoh} satellite data. 
We appreciate the BASJE group and members of Universidad Mayor de San Andr\'{e}s for management and maintenance of the Bolivia solar neutron telescope and the neutron monitor.
We also thank Prof. John Hearnshaw for reading this manuscript.
We wish to thank the referee for evaluating this paper and for helping us to clarify several arguments.

\end{document}